\documentclass[12pt,a4paper,oneside,onecolumn]{article}
\usepackage[a4paper]{geometry}
\usepackage{lipsum}
\usepackage{times}
\usepackage{amsmath}
\usepackage{amssymb}
\usepackage{float}
\usepackage[english]{babel}
\usepackage[utf8]{inputenc}
\usepackage{amsmath}
\usepackage{bbm}
\usepackage[colorinlistoftodos]{todonotes}
\usepackage[labelfont=bf]{caption}
\emergencystretch=\maxdimen
\usepackage{tocloft}
\usepackage{booktabs}
\usepackage[titletoc]{appendix}
\usepackage{tcolorbox}

\usepackage{csquotes}
\usepackage[style=authoryear-ibid,backend=biber]{biblatex}
\usepackage{graphicx}
\usepackage{lscape}

\title{\vspace{-2.0cm}Evaluation of Dynamic Cointegration-Based Pairs Trading Strategy in the Cryptocurrency Market} 
\author{
    \textsc{Masood Tadi}\\[1ex] \normalsize CY Tech, CY Cergy Paris University, Cergy, France \\
    \normalsize \href{mailto:tadimasood@cy-tech.fr}{tadimasood@cy-tech.fr} 
    \and
    \textsc{Irina Kortchemski}\\[1ex] \normalsize CY Tech, CY Cergy Paris University, Cergy, France \\ 
    \normalsize \href{mailto:ik@cy-tech.fr}{ik@cy-tech.fr}
    }
\date{}
\usepackage{hyperref}
\hypersetup{
    colorlinks=true, 
    citecolor=blue,
    linktoc=all,     
    linkcolor=blue,  
    urlcolor=blue} 
\begin{document}
    \maketitle
	\setcounter{secnumdepth}{5}	
	\setcounter{tocdepth}{4}
	\pagenumbering{roman}
    \pagenumbering{arabic}
    
    \section*{Abstract}
        \textbf{Purpose} \textemdash \, This research aims to demonstrate a dynamic cointegration-based pairs trading strategy, including an optimal look-back window framework in the cryptocurrency market, and evaluate its return and risk by applying three different scenarios.\par
        \textbf{Design/methodology/approach} \textemdash \, We employ the Engle-Granger methodology, the Kapetanios-Snell-Shin (KSS) test, and the Johansen test as cointegration tests in different scenarios. We calibrate the mean-reversion speed of the Ornstein-Uhlenbeck process to obtain the half-life used for the asset selection phase and look-back window estimation.\par 
        \textbf{Findings} \textemdash  \, By considering the main limitations in the market microstructure, our strategy exceeds the naive buy-and-hold approach in the Bitmex exchange. Another significant finding is that we implement a numerous collection of cryptocurrency coins to formulate the model's spread, which improves the risk-adjusted profitability of the pairs trading strategy. Besides, the strategy's maximum drawdown level is reasonably low, which makes it useful to be deployed. The results also indicate that a class of coins has better potential arbitrage opportunities than others. \par 
        \textbf{Originality/value} \textemdash \, This research has some noticeable advantages, making it stand out from similar studies in the cryptocurrency market. First is the accuracy of data in which minute-binned data create the signals in the formation period.  Besides, to backtest the strategy during the trading period, we simulate the trading signals using best bid/ask quotes and market trades. We exclusively take the order execution into account when the asset size is already available at its quoted price (with one or more period gaps after signal generation). This action makes the backtesting much more realistic.\par 
        \textbf{Keywords} \textemdash \, Arbitrage opportunity, Pairs trading strategy, Basket trading, Cointegration, Mean reversion, Cryptocurrency market\par 
        \textbf{Paper type} \textemdash \, Research paper
    \newpage
    
    \section{Introduction}\label{ch 1}
        A cryptocurrency is volatile digital security designed to operate as a tool of exchange that uses secure cryptography \parencite{gandal2016can}. Bitcoin is the first cryptocurrency that is issued in 2009. After the bitcoin announcement, a lot of alternative cryptocurrencies (altcoins) have been created. The total market capitalization of the cryptocurrency market is more than $341$ billion dollars on 30/09/2020 (See Coinmarketcap.com). Deploying algorithmic trading strategies is developing over time in the Cryptocurrency market, accelerating tradings to maximize profits.\par
        One of the most well-known strategies among different algorithmic trading methods is the statistical arbitrage strategy. Statistical arbitrage is a profitable situation stemming from pricing inefficiencies among financial markets. Statistical arbitrage is not a real arbitrage opportunity, but it is merely possible to obtain profit applying past statistics \parencite{aldridge2013high}. In fact, there are two different potential arbitrage opportunities in the cryptocurrency market; the exchange to exchange arbitrage and the statistical arbitrage. The exchange to exchange arbitrage has a potential profit, but it is quite risky, and there are many challenges to deploy it. Instead, the statistical arbitrage opportunities have the same potential profits without the same risks as the former one \parencite{pritchard2018digital}.\par
        Statistical arbitrage strategies are often deployed based on mean reversion property, but they can also be designed using other factors. Pairs trading is the commonly recognized statistical arbitrage strategy that involves identifying pairs of securities whose prices tend to move mutually. Whenever the relationship between financial securities behaves abnormally, the pair would be traded. Then the open positions will be closed when the unusual behavior of pairs reverts to their normal mode \parencite{vidyamurthy2004pairs}.\par
        According to \textcite{krauss2017statistical}, pairs trading strategy is a two-step process. The first step, which is called the formation period, attempts to find two or more securities whose prices move together historically. In the second step, which is the trading period, we seek abnormalities with their price movement to profit from statistical arbitrage opportunities. \par 
        There are two general approaches to find appropriate pairs of assets in the formation period: the heuristic approach and the statistical approach. The heuristic approach, which is regularly called the distance approach, is more straightforward than the statistical approach, which the latter is based on the cointegration concept. In the trading period, we can combine our strategy with different mathematical tools such as stochastic processes, stochastic control, machine learning, and other methods to improve the results \parencite{krauss2017statistical}. \par
        Compared to the leading financial markets, such as the stock market and fixed income market, limited research has been conducted on statistical arbitrage strategies in the cryptocurrency market. In the next section, we review some papers concerning these strategies frequently studied in the cryptocurrency market and explain some of their practical weaknesses and drawbacks.\par
        
    \section{Literature Review}\label{ch 2}
    	The distance approach is the basic researched framework introduced by \textcite{gatev2006pairs}. This method is based on the minimum squared distance of the normalized price of assets. The distance between normalized prices is called spread. In order to construct a measurable scale, assets should become normal first. To this aim, asset prices are divided into their initial value, and then the spread is obtained by taking the difference between normalized prices. In the formation period, the top first pairs with the minimum historic sum of squared distances between normalized prices are considered in a subsequent trading period \parencite{gatev2006pairs}. \par 
        According to \textcite{perlin2009evaluation}, the price series can be normalized first based on their historical mean and historical volatility, and then the spread of each pair can be constructed in the same way. Finally, in the trading period, the pairs arranged in their ascending orders can be picked using the top first pairs of the list for pairs formation. Simple non-parametric threshold rules are used to trigger trading signals. This threshold and can be two historical standard deviations of normalized spread price. \par
		The main findings establish distance pairs trading as profitable across different markets, asset classes, and time frames. \textcite{driaunys2014algorithm} studied a pairs trading strategy at the natural gas futures market based on the distance approach. In their research, pairs of two futures contracts of the same underlying asset are selected, i.e., natural gas with different maturities, which are the most liquid, close to expiration, and therefore they are correlated contracts of natural gas futures. The contract with bid prices is shorted through the trading period, Whenever the distance of the given $ d $ is reached. Besides, the long position of the contract with the asking price is taken. They backtest their model with different moving windows and different thresholds and realize that the higher values of $d$ works better and generate fewer trades and make the model more stable. \par
	    \textcite{driaunys2014algorithm}'s research has some weaknesses in practice. First, the assumption of no transaction cost makes their result unreliable. Second, they did not calculate the risk-adjusted return of different scenarios. So, the performance of the research's strategy is questionable. Furthermore, they did not compare their method with a naive strategy, i.e., buy-and-hold strategy, and besides, the required investment to deploy the strategy is not determined.\par
		Other researches are based on the statistical approach. This approach identifies two or more time-series combined to form a long-term equilibrium relationship, although the time series themselves may have a non-stationary trend. There are some research papers based on the cointegration concept in the cryptocurrency market. \textcite{van2018cointegration} selected a set of cryptocurrency coins and split them into four main sectors that depend on the coins' fundamental features.  Then, by examining unit-root tests, they achieved a cointegration relationship among coins in each group. Finally,  they concluded that implementing a pairs trading strategy could be profitable due to arbitrage opportunities.  Nevertheless, ignoring the transaction costs and having a bias selection make this paper's profitability results unreliable.\par
		\textcite{leung2019constructing} constructed cointegrated cryptocurrency pairs using three different unit-root tests.  Using the daily prices of four cryptocurrency coins from Coinbase exchange, they introduced the ordinary least squares model to build a cointegrated combination of coins and set the p-values of the estimated coefficients less than 1\%. The authors backtested the strategy with five different entry/exit threshold levels. They realized that with the threshold set at $1.5$ standard deviation, the strategy is optimal. Furthermore, they incorporated the stop-loss exit and trailing stop-loss exit possibilities, which lowered the profit return.\par
        \textcite{kakushadze2019altcoin} proposed the momentum factor statistical arbitrage methodology based on a dollar-neutral mean-reversion strategy. Their method is to short a fixed level of Bitcoin and keep it throughout the trading period and long multiple altcoins, which varies every day depending on their momentum values. They realized that low liquid altcoins have a better mean-reverting feature. Furthermore, they found that if yesterday's momentum of an altcoin is positive, it is expected to trade it higher today and vise versa.\par
        \textcite{pritchard2018digital} built the strategy based on statistical tests as well as technical analysis indicators. He applied different statistical tests such as the augmented Dickey-Fuller test, the Hurst exponent, and the Johansen test. He achieved that each coin is not mean-reverting, but a linear mean-reverting strategy can be implemented by constructing the normalized deviation of price from its moving average. In his study, the volume is considered the most critical barrier to utilize arbitrage opportunities in the cryptocurrency market.\par
        
	\section{Market Structure}\label{ch 3}
        Numerous cryptocurrency exchanges allow their customers to trade cryptocurrencies against other assets, such as conventional fiat money or other digital currencies. BitMEX is one of the most well-known cryptocurrency exchanges with a Peer-to-Peer Trading Platform, which offers leveraged contracts bought and sold in Bitcoin. This exchange only handles Bitcoin. All profit and loss are in Bitcoin, even if altcoins are traded. BitMEX does not handle fiat currencies. Therefore, in order to calculate profit and loss, we denominate all coins by Bitcoin. BitMEX offers Futures and Perpetual Contracts and allows trading with a high amount of leverage depending on the type of contracts. BitMEX offers perpetual swaps that have an inverse or Quanto payout. It also offers futures contracts that have an inverse, Quanto, or linear payout. See \textcite{Bitmex} for more details.
        \begin{table}[H]
            \centering
            \begin{tabular}{@{}cl@{}}
                \toprule
                \multicolumn{1}{l}{\textbf{Symbol}} &
                \multicolumn{1}{l}{\textbf{Description}} \\ \midrule
                XBT/USD & Bitcoin/US Dollar (Inverse) Perpetual Swap \\
                ETH/USD & Ethereum Quanto Perpetual Contract \\
                ETH/XBT & Ethereum/Bitcoin  Futures Contract \\
                EOS/XBT & EOS Token/Bitcoin  Futures Contract \\
                LTC/XBT & Litecoin/Bitcoin  Futures Contract \\
                XRP/XBT & Ripple/Bitcoin  Futures Contract \\
                BCH/XBT & Bitcoin Cash/Bitcoin  Futures Contract \\
                TRX/XBT & Tron/Bitcoin  Futures Contract \\
                ADA/XBT & Cardano/Bitcoin  Future Contract \\ \bottomrule
            \end{tabular}
            \caption{BitMEX contracts used in the research}
            \label{table 1}
        \end{table}
        \raggedbottom
        BitMEX uses the following industry-standard month codes to name its Futures coins. The month code implies the delivery month of the Futures contract.
        \begin{table}[H]
        \centering
        \begin{tabular}{@{}llll@{}}
        \toprule
        Month Code & Month & Month Code & Month \\ \midrule
        F & Jan & N & Jul \\
        G & Feb & Q & Aug \\
        H & Mar & U & Sep \\
        J & Apr & V & Oct \\
        K & May & X & Nov \\
        M & Jun & Z & Dec \\ \bottomrule
        \end{tabular}
        \caption{Futures Expiration Month Codes on BitMEX }
        \label{table 2}
        \end{table}
        Each Futures contract is traded in three months period. In this research, the delivery date of futures contracts are \textit{Z18}, \textit{H19}, \textit{M19}, and \textit{U19}. The letters are the expiration month code, and the numbers show the expiration year. For instance, Litecoin has four different delivery dates: LTCZ18, LTCH19, LTCM19, and LTCU19.\par
        \begin{figure}[H]
    		\includegraphics[scale=0.9]{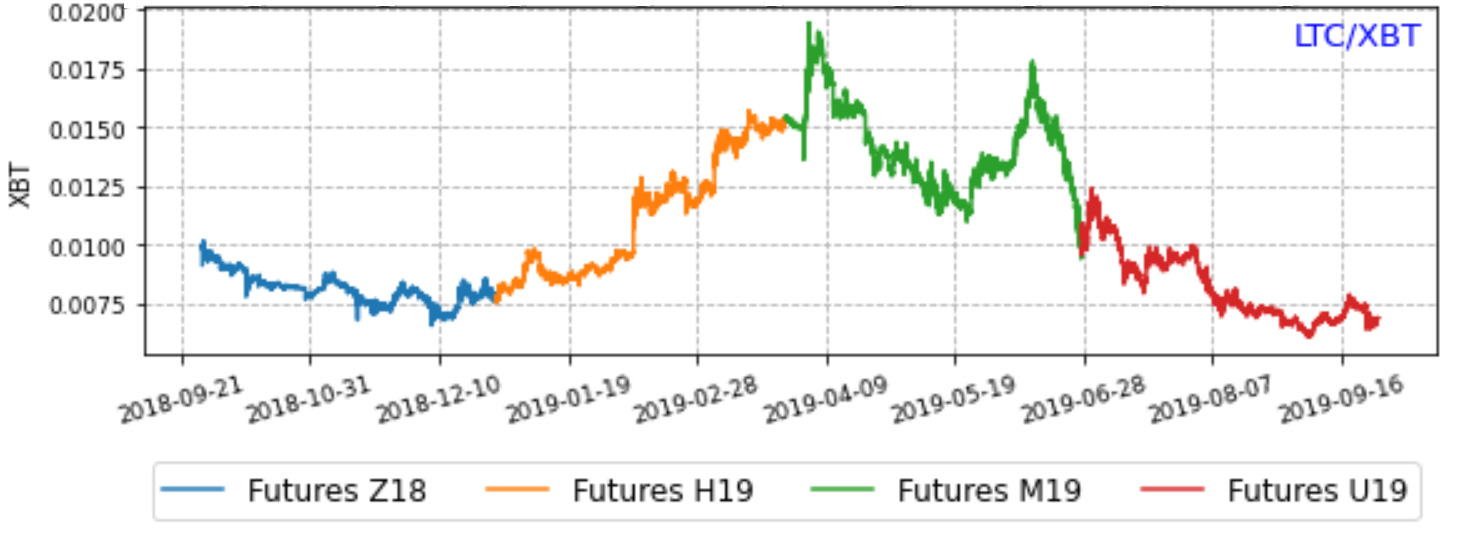}
    		\centering
    		\caption {LTC/XBT futures contract price during the study period}
    		\label{fig 3.1}
	    \end{figure}
        Bitcoin perpetual swap on BitMEX (XBT/USD) is a derivative product whose underlying asset is Bitcoin in the US dollar. Since Bitcoin's price in the US dollar is different in each cryptocurrency exchange, Bitmex identifies an index named BXBT, which calculates a weighted average Bitcoin price in the US dollar using several exchanges data sources. Each Bitcoin perpetual swap contract is worth 1 USD of Bitcoin. Note that both underlying and swap contracts are quoted in USD, but margin and P\&L are denominated in Bitcoin. It means that opening a long position of XBT/USD is equal to open a short position of the US dollar, which is denominated in Bitcoin. See \textcite{Bitmex} for more details.\par
	    \begin{figure}[H]
    		\includegraphics[scale=0.62]{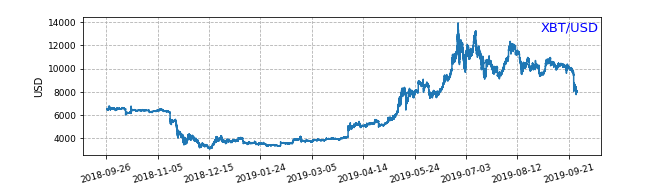}
    		\centering
    		\caption {XBT/USD perpetual swap price during the study period}
    		\label{fig 3.2}
	    \end{figure}
        When we long or short a contract, there are two fundamental execution options: placing the market order or the limit order. Market orders are executed immediately at the market price of an asset. Conversely, in limit orders, we set a specific price that we desire to trade. Hence there is no guarantee to have executed trades. The main difference between these two orders is their transaction fees. Taker fees (market orders fees) typically have a higher charge than maker fees (limit orders). In BitMEX exchange, maker fees for Perpetual and Futures contracts are negative ($-0.0250\%$), which means that by executing a limit order, we are not charged and can earn profit due to the negative transaction fees. Taker fee is $0.0750\%$ for all contracts. Besides, in order to match the transaction price of a perpetual contract with its underlying coin price, there is another fee, called funding fee which is exchanged between long and short positions every 8 hours. Funding fee is $\pm 0.0100\%$ for XBT/USD and ETH/USD perpetual contracts.\par 
        There are also other execution orders in BitMEX such as stop-loss orders, trailing stop-loss orders, and take profit orders. The stop-loss order and trailing stop-loss order place a market/limit order to close a position to restrict an investor's loss on a cryptocurrency position. On the other hand, the take profit order places a market/limit order to exit the trade to maximize investor's profit in unstable or bubble periods. As \textcite{geuder2019cryptocurrencies} studied Bitcoin price dynamics, the cryptocurrency coins (Bitcoin, in their paper) have remarkable bubbles in specific periods. However, in other periods, their used methodologies did not identify any bubble behavior. To overcome high risky periods of bubble behavior of cryptocurrency coins, we can use these kinds of orders to protect our investment and decrease the maximum drawdown of the strategy.

	\section{Data Source}\label{ch 4}
	    To implement the statistical arbitrage strategy, we obtain historical minute-binned trading data of cryptocurrency coins using the BitMEX application programmable interface (API) from September 27, 2018, to October 2, 2019. The data consist of the trade's close price and the volume of each coin as well as best bid/ask quotes. We used the data in two steps. First, for each coin, the minute-binned close price is used to calculate cointegration tests, making a spread, calibrating half-life of the spread, and creating trading signals of the strategy. Second, to backtest the strategy, we simulate the orders by updating best bid/ask quotes and comparing them with market trades. \par
\begin{figure}[H]
    		\includegraphics[scale=0.79]{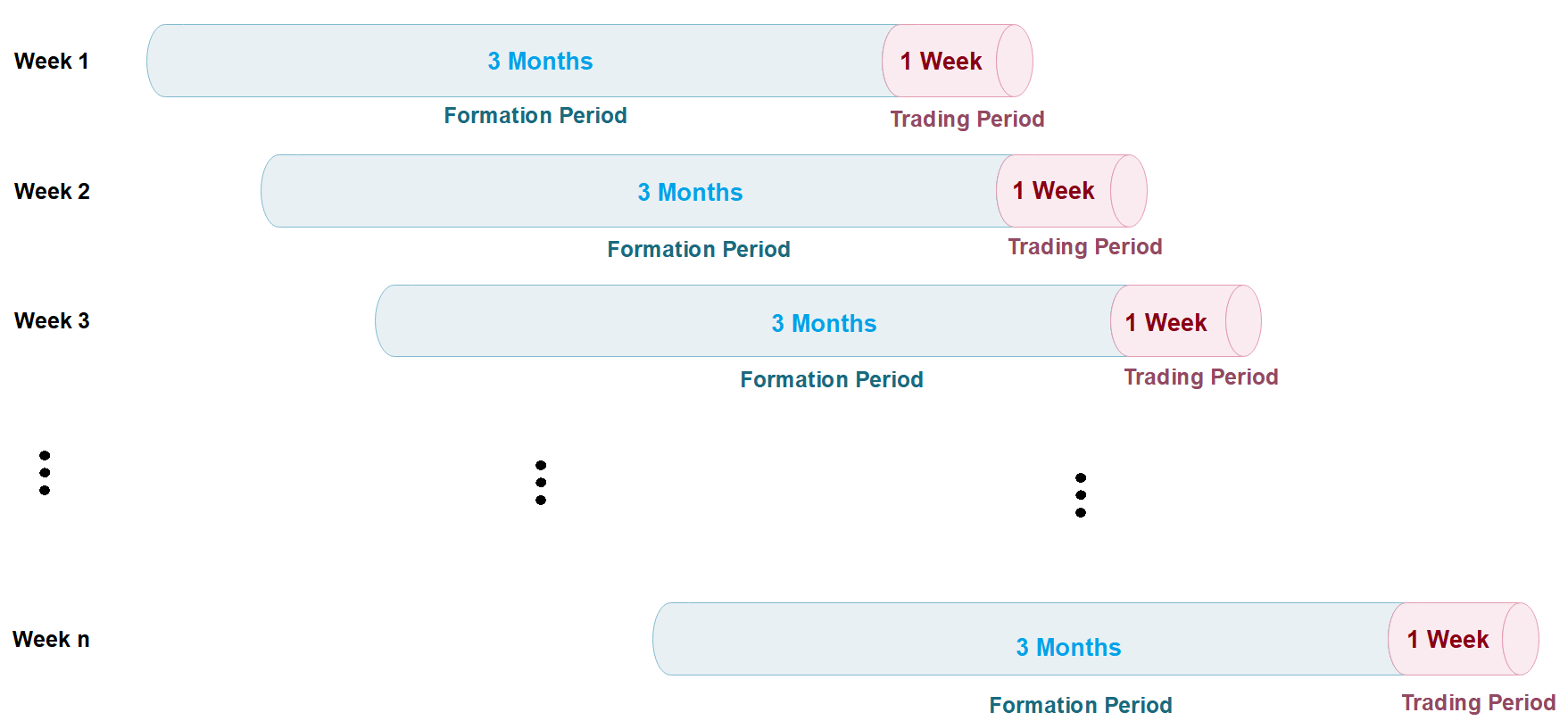}
    		\centering
    		\caption {Dynamic Formation-Trading Periods}
    		\label{fig 3.3}
	    \end{figure}

	    The data splitting step is dynamic. In order to create trading signals, we need to specify a formation period to estimate our parameters. The formation period includes three-month price data of coins. After calibrating the parameters, we start trading for one week, corresponding to the estimated parameters. The first trading week starts on 26/12/2018. As shown in figure \ref{fig 3.3}, after the first trading week, we update the formation period and move the time horizon ahead. There are thirty-nine trading weeks in which the last one starts from 18/09/2019. \par
	    The strategy is programmed by Python3 language and its helpful libraries such as Pandas, Numpy, Multiprocessing, Matplotlib, Statsmodels, and Arch. We also used R packages to perform nonlinear cointegration tests. The data is stored on Amazon Web Service (AWS) cloud server. To optimize the running time and facilitate programming, we worked on JupyterLab in the Amazon SageMaker platform.\par

    \section{Theoretical Framework of the Strategy}	\label{ch 5}
        This section explains the theoretical concepts, such as statistical tests and calibration procedures required to develop our methodology.
        \subsection{Non-Stationary Processes and Unit-Root Test}
            Generally, financial variables are regularly non-stationary, which means that the data parameters change over time. This phenomenon makes us utilize cointegration property. If we have a collection of non-stationary time series with the same order of integration, denoted $I(d)$, and if a linear combination of them is stationary, then the collection is cointegrated. Random walks are well-known examples of non-stationary processes that can have drift and deterministic trend.\par
            In a statistical arbitrage strategy, we create a combination of two or more assets in which some of them are on the buy-side, and some others are on the sell-side. This combination is called the assets spread.  The methodology of formulating that spread is discussed in section \ref{ch 6}. Suppose that $S_t$ is the spread time series. Using unit-root tests, we check if $S_t$ is a non-stationary process. In this research, the spread process is formulated such that the unit-root test is viewed without drift or a trend as it should revert to zero as follows
            \begin{equation}\label{eq 4.1}
                S_t = \phi_1S_{t-1} + \epsilon_t ,
            \end{equation}
            where $\epsilon_t$ indicates the error term. Consider the null hypothesis $H_0 : \phi_1 = 1$ versus the alternative hypothesis $H_1 : \phi_1 < 1$. In this case, if the series $S_t$ is stationary, it tends to return to a zero mean. The Dickey–Fuller (DF) test statistic is the t ratio estimation of the least-squares of $\phi_1$ \parencite{tsay2010analysis}. In general, it is  easier to test a null hypothesis that a coefficient is equal to zero, $S_{t-1}$ is subtracted from both sides of equation \ref{eq 4.1}, to obtain
            \begin{equation}\label{eq 4.2}
                \Delta S_t = S_t - S_{t-1}= \gamma_1S_{t-1} + \epsilon_t,
            \end{equation}
            where $\gamma_1=\phi_1-1$. This model can be estimated and testing for a unit-root is equivalent to testing $\gamma_1=0$. The corresponding null and alternative hypotheses are now $H_0 : \gamma_1 = 0$ against $H_1 : \gamma_1 < 0$. Now we extend the price time series to lags more than one such that
            \begin{equation}\label{eq 4.3}
                S_t = \phi_1S_{t-1} +\phi_2S_{t-2}+\cdots+\phi_pS_{t-p}+ \epsilon_t,
            \end{equation}
            This series is stationary if the roots of the polynomial $m^{p}-m^{p-1}a_{1}-m^{p-2}a_{2}-\cdots -a_{p}$ are all less than one in absolute value. In this case, One is a root of this polynomial if $1 -\phi_1 -\phi_2 - \cdots -\phi_k = 0 \implies \sum_{i=1}^p\phi_i=1$. Note that when $p = 2$  we can re-write an the price process as follows
            \begin{equation}\label{eq 4.4}
                \begin{split}
                    S_t &= \phi_1S_{t-1} +\phi_2S_{t-2}+\epsilon_t\\
                    &= \phi_1S_{t-1} +\phi_2S_{t-1}-\phi_2S_{t-1}+\phi_2S_{t-2}+\epsilon_t\\
                    &=(\phi_1+\phi_2)S_{t-1}+\gamma\Delta S_{t-1}+\epsilon_t
                \end{split}
            \end{equation}
            where $\gamma=f(\phi_2)$. In general we can rewrite price process as follows
            \begin{equation}\label{eq 4.5}
                S_t = \left(\sum_{i=1}^p\phi_i\right)S_{t-1} + \gamma_1\Delta S_{t-1}+\cdots+\gamma_{p-1}\Delta S_{t-p+1}+\epsilon_t
            \end{equation}
            where $\gamma_i$ is $f(\phi_{i+1}, \cdots, \phi_p)$. By defining  $\beta = \sum_{i=1}^p\phi_i$, the equation is
            \begin{equation}\label{eq 4.6}
                S_t = \beta S_{t-1} + \gamma_1\Delta S_{t-1}+\cdots+\gamma_{p-1}\Delta S_{t-p+1}+\epsilon_t
            \end{equation}\par
            To verify the existence of a unit-root in a spread process, one may perform the test $H_0 : \beta = 1$ vs. $H_a : \beta < 1$. This test is the augmented Dickey–Fuller (ADF) unit-root test \parencite{dickey1979distribution}. If coefficient $\text{p-value} < \alpha$, we reject unit-root non-stationary time series with $(1-\alpha)\%$ of statistical confidence and if coefficient $\text{p-value}\geq\alpha$, we cannot reject unit-root non stationary time series with $(1-\alpha)\%$ of statistical confidence \parencite{tsay2010analysis}.\par
            The primary point in the Dickey-Fuller test is that we assume the spread of the pairs trading strategy is a linear Auto-regressive time series. We can extend equation \ref{eq 4.3} to Piece-wise nonlinear time series which is called Threshold Auto-Regressive (TAR) model such that
            \begin{equation}\label{eq 4.7}
                S_t = 
                \begin{cases} 
                    \phi_{11}S_{t-1} +\phi_{12}S_{t-2}+\cdots+\phi_{1p}S_{t-p}+ \epsilon_t, & z_{t-d}<=c \\
                    \phi_{21}S_{t-1} +\phi_{22}S_{t-2}+\cdots+\phi_{2p}S_{t-p}+ \epsilon_t & z_{t-d}>c
                \end{cases}
            \end{equation}
            where $z_t$ is the transition variable,  $d$ is the delay of the transition variable, $c$ is the threshold of the model, and $\epsilon_t$ indicates the error term. Equation \ref{eq 4.7} is a two regime TAR model but theoretically, the number of regimes can be more than two. If $z_{t-d}=S_{t-d}$, then The switch from one regime to another regime depends only on the past values of the $S_t$. In this case the model is named Self-Exciting Threshold Auto-Regressive (SETAR) model. \textcite{enders2001cointegration} proposed a three-step method to identify whether there exists cointegration among time series in which their residuals $(S_t)$ are formed by the SETAR model. They developed their method both when $c$ is known ($c = 0$) and when it is unknown. We also demonstrated that their test has greater power compared to the Engle-Granger test when $z_{t-d} = \Delta S_{t-d}$. Now, We generate three regime TAR model with one inner regime and two outer regimes as follows
            \begin{equation}\label{eq 4.8}
                S_t = 
                \begin{cases} 
                    \phi_{11}S_{t-1} +\phi_{12}S_{t-2}+\cdots+\phi_{1p}S_{t-p}+ \epsilon_t, & z_{t-d}<c \\
                    \phi_{21}S_{t-1} +\phi_{22}S_{t-2}+\cdots+\phi_{2p}S_{t-p}+ \epsilon_t & z_{t-d}= c\\
                    \phi_{31}S_{t-1} +\phi_{32}S_{t-2}+\cdots+\phi_{3p}S_{t-p}+ \epsilon_t, & z_{t-d}>c \\
                \end{cases}
            \end{equation}
            \textcite{chan1985multiple} proved that the inner regime of equation \ref{eq 4.8} can be non-stationary while the entire process can be globally stationary. Hence, a unit root in the inner regime does not influence the stationarity of the entire process and we can replace the inner regime by $S_{t-1} + \epsilon_t$. Assume that $\phi_{1i} = \phi_{3i}$ for $i = 1, \cdots, p$ (which means that the SETAR model is symmetric), we deduce
            \begin{equation}\label{eq 4.9}
                S_t = 
                \begin{cases} 
                    S_{t-1}+ \epsilon_t & z_{t-d}= c\\
                    \phi_{11}S_{t-1}+\phi_{12}S_{t-2}+\cdots+\phi_{1p}S_{t-p}+ \epsilon_t, & z_{t-d}\neq c 
                \end{cases}
            \end{equation}

            By aassuming $z_{t-d}=S_{t-d}$, and $d=1$, the equation \ref{eq 4.7} can be written in the form of indicator function as follows
            \begin{equation}\label{eq 4.10}
                \begin{split}
                    S_t &= S_{t-1}\left(1 -\mathbbm{1}_{\{S_{t-1}\neq c\}}\right)+\left(\sum_{i=1}^p\phi_{1i}\,S_{t-i}\right)\mathbbm{1}_{\{S_{t-1}\neq c\}}+ \epsilon_t\\  
                    &=S_{t-1}+\sum_{i=1}^p\left(\gamma_{1i}\,\mathbbm{1}_{\{S_{t-1}\neq c\}}\,S_{t-i}\right)+ \epsilon_t 
                \end{split}
            \end{equation}
            where $\gamma_{1i}= \phi_{1i} - 1$. The transition from one regime to another in the TAR and SETAR models are discontinuous. Hence, when $c$ is unknown, the analytical form to obtain an estimator for $c$ cannot be obtained. The overcome this issue, the $\mathbbm{1(.)}$ is replaced by a continuous function to have a higher level of flexibility in the model. So the Smooth Transition Auto-regressive (STAR) model is
            \begin{equation}\label{eq 4.11}
                S_t = S_{t-1}+\sum_{i=1}^p\left(\gamma_{1i}\,G(S_{t-1};\theta,c)\,S_{t-i}\right)+ \epsilon_t 
            \end{equation}
            where
            \begin{equation}\label{eq 4.12}
                G(S_{t-1};\theta,c)=1-e^{-\theta\left(S_{t-1}-c\right)^2}.
            \end{equation}
            Equations \ref{eq 4.11} and \ref{eq 4.12} generate an Exponential Smooth Transition Auto-Regressive (ESTAR) model. In this model, $0\leq G(S_{t-1};\theta,c)\leq1$. For simplicity, in the case of $p=1$, and $c=0$, ESTAR(1) is considered as  
            \begin{equation}\label{eq 4.13}
                S_t =\left(1+\gamma_{1}G(S_{t-1};\theta)\right)\,S_{t-1}+ \epsilon_t 
            \end{equation}
            which can be demonstrated as
            \begin{equation}\label{eq 4.14}
                \Delta S_t =\gamma_{1}\left(1-e^{-\theta S_{t-1}^2}\right)\,S_{t-1}+ \epsilon_t 
            \end{equation}
            \textcite{kapetanios2003testing} (hereafter KSS) presented a test for nonstationarity with the alternative hypothesis of nonlinear stationarity. They showed that $S_t$ is globally stationary when $\theta>0$ and $-2<\gamma<0$. The null hypothesis and its alternative hypothesis are $H_0: \theta=0$ and $H_1:\theta>0$. The problem is that $\gamma$ is not identified under the null hypothesis. In order to overcome this problem, they suggested to take first-order Taylor approximation of the $G(S_{t-1};\theta)$. Suppose that $G(S_{t-1};\theta) = 1-f(S_{t-1};\theta)$. The Taylor approximation of $f(\theta)$ at $\theta = 0$ is
            \begin{equation}\label{eq 4.15}
                \begin{split}
                    f(S_{t-1};\theta) 
                    &= f(S_{t-1};\theta|\theta=0)+\frac{\partial f(f(S_{t-1};\theta|\theta=0)}{\partial\theta}\left(\theta-0\right)+C\\
                    &=1-\theta\left(S_{t-1}\right)^2+C
                \end{split}
            \end{equation}
            where C is the further orders of the approximation. Now, the first order Taylor approximation of $G(S_{t-1};\theta)$ at $\theta = 0$ is $\theta\left(S_{t-1}\right)^2$. By replacing this expression into equation \ref{eq 4.14} we obtain
            \begin{equation}\label{eq 4.16}
                \Delta S_t = \gamma_{1}\theta \left(S_{t-1}\right)^2\,S_{t-1}+ \epsilon^*_t =\delta \left(S_{t-1}\right)^3+ \epsilon^*_t
            \end{equation}
            where $\delta = \gamma_{1}\theta$ and $\epsilon^*_t = -(\gamma_1y_{t-1})C+\epsilon_{t}$.
            The null hypothesis and alternative hypothesis of KSS test are now $H_0: \delta=0$ and $H_1:\delta<0$. See \textcite{kapetanios2006testing}, \textcite{kapetanios2003testing}, and \textcite{patterson2012unit} for more details about KSS test characteristics.\par
            Finally, to test the cointegration test among multiple assets, we need to utilize the \textcite{johansen1991estimation} test, a procedure for testing the cointegration of several non-stationary time series. This test permits more than one cointegration relationship, so it is more generally applicable than the \textcite{engle1987co} test based on the Dickey-Fuller (or the augmented) test for unit-roots in the residuals from a single (estimated) cointegration relationship. Like a unit-root test, suppose that we have an auto-regressive process with order $p$, which is now in vector shape, named VER($p$) model. We assume that there is no trend term
            \begin{equation}\label{eq 4.17}
                S_{t}= \Pi_{0} + \Pi _{1}S_{t-1}+\cdots +\Pi_{p}S_{t-p}+\varepsilon_{t},\quad t=1,\dots ,T.
            \end{equation}
            Like the augmented Dickey-Fuller test, we rewrite the spread process by differencing the series we obtain
            \begin{equation}\label{eq 4.18}
                \Delta S_{{t}}=\Pi_0+ \Pi S_{{t-1}}-\Gamma _{{1}}\Delta S_{{t-1}}-\cdots -\Gamma _{{p-1}}\Delta S_{{t-p+1}}+\varepsilon _{{t}},\quad t=1,\cdots ,T
            \end{equation}
            where $ \Gamma _{i}=\left(\Pi _{{i+1}}+\cdots +\Pi _{p}\right),\quad i=1,\dots ,p-1$, and $\Pi =\Pi _{{1}}+\cdots +\Pi _{{p}}-I$. $\Delta S_{t}=S_{t}-S_{t-1}$ is the difference operator, $\Pi$ and $\Gamma_i$ are the coefficient matrix for the lags. Cointegration occurs when the matrix $\Pi = 0$. We decompose the eigenvalue of $\Pi$. Suppose that the rank of $\Pi$ is equal to $r$. The Johansen test's null hypothesis is $H_0: r=0$ indicates no cointegration at all. A rank $H_1: r>0$ means a cointegration relationship within two or perhaps more time series. In this test, we sequentially test whether $r = 0, 1, p-1$, where $p$ is the number of time series under test. Finally we find coefficients of a linear combination of time series to produce a stationary portfolio of time series.\par
        
        \subsection{Mean Reverting Processes and Half-Life Calibration}
            Mean reverting processes are widely observed in the financial markets. Toward mean reverting processes, as opposed to the trend following time series, we assume that a combination of assets tends to return to its mean level over time.  A mean-reverting time series can be interpreted by \textcite{uhlenbeck1930theory} model or \textcite{vasicek1977equilibrium} model, a particular case of the \textcite{hull1990pricing} model with constant volatility. It is also the continuous-time equivalent of the discrete-time Auto-Regressive process with order $1$. The Ornstein-Uhlenbeck model specifies the stochastic differential equation as follows
		    \begin{equation}\label{eq 4.19}
			    dS_t = \theta (\mu - S_t)\,dt +\sigma\,dW_t
		    \end{equation}
		    where $S_0=s_0$ is a known value, $\theta>0$, $\mu>0$. $ W_t $ denotes the Wiener process. In this model, $\mu$ is the long term spread mean, $\theta$ is the mean reversion speed, and $\sigma$ is the spread's instantaneous volatility. We rearrange the equation \ref{eq 4.19}, multiply the integrating factor of $e^{\theta t}$  and integrate both sides knowing $e^{\theta t}\,dS_t + e^{\theta t}\theta S_t\, dt=d(e^{\theta t}S_t)$ as follows
	        \begin{equation}\label{eq 4.20}
			    \int^{t=T}_{t=s} d(e^{\theta t}S_t)\,dt = \int^T_s e^{\theta t}\theta \mu\, dt +\int^T_s e^{\theta t}\sigma\,dW_t 
		      \end{equation}
		      So we obtain
            \begin{equation}\label{eq 4.21}
		        e^{\theta T}S_T - e^{\theta s}S_s = \int^T_s e^{\theta t}\theta \mu\,dt +\int^T_s e^{\theta t}\sigma\,dW_t.
		    \end{equation}
		    Solving for $ S_T $ we get
		    \begin{equation}\label{eq 4.22}
		    	S_T=e^{-\theta (T-s)} S_s+\mu(1-e^{-\theta (T-s)})+ \sigma e^{-\theta T}\int^T_s e^{\theta t}\,dW_t
	    	\end{equation}		    
		    where the expected value of $S_t$ is 
		    \begin{equation}\label{eq 4.23}
		        \mathbb{E}[S_{t}|s_0]=s_{0}e^{{-\theta t}}+\mu(1-e^{{-\theta t}}).
		    \end{equation}
		    Concerning the equation \ref{eq 4.23}, we can define its half-life. Half-life $(t_{1/2})$ is the expected time required for any quantity to decrease to half of its initial value.  Consequently, the half-life of $S_t$ is the time that the expected value of $S_t$ reach the average value between $S_0$ and $\mu$ as follows
		    \begin{equation}\label{eq 4.24}
		       \mathbb{E}[S_{t_{1/2}}|s_0] - \mu =\frac{ s_0 -\mu}{2}.
		    \end{equation}
		    By rewriting left-hand-side of the equation \ref{eq 4.24} with respect to the equation \ref{eq 4.23}, we deduce
		    \begin{equation}\label{eq 4.25}
		        \begin{split}
		            \mathbb{E}[S_{t_{1/2}}|s_0]-\mu &=s_{0}e^{{-\theta t_{1/2}}}+\mu(1-e^{{-\theta t_{1/2}}}) - \mu\\ &= e^{-\theta t_{1/2}}(s_0 - \mu)
		        \end{split}
		    \end{equation}
		    Solving it with the right-hand-side of the equation \ref{eq 4.24} gives the half-life as follows
		    \begin{equation}\label{eq 4.26}
		       e^{-\theta t_{1/2}}(s_0 - \mu) = \frac{s_0 - \mu}{2}\quad \Rightarrow \quad t_{1/2}=\frac{\ln{2}}{\theta}
		    \end{equation}
	        To calibrate the parameter $\theta$, we generally have two approaches, i.e., the Least square and Maximum Likelihood Estimation \parencite{smith2016simulation}. The most straightforward approach is converting the stochastic differential equation to a finite difference equation and rearranging parts to the Ordinary Least Squares equation. The finite-difference formula for the Ornstein-Uhlenbeck process is as follows
            \begin{equation}\label{eq 4.27}
                \begin{split}
                    S_t - S_{t-1} &= \theta (\mu - S_{t-1}) \Delta t + \sigma \sqrt{\Delta t} W_{t-1}\\ &= \theta\mu\Delta t - \theta S_{t-1}\Delta t +  \sigma \sqrt{\Delta t} W_{t-1}.   
                \end{split}
            \end{equation}
            Matching with a simple regression formula $y = a + bx + \epsilon$ we can equate $y = S_t - S_{t-1}$, $x = S_{t-1}$, $a = \theta\mu\Delta t$, $b = -\theta\Delta t$, and $\epsilon = \sigma \sqrt{\Delta t} W_{t-1}$ and consequently obtain
            \begin{equation}\label{eq 4.28}
                \theta = -\frac{b}{\Delta t},
            \end{equation}
            So, regression of $S_{t-1}$ against $S_t - S_{t-1}$ gives estimation of parameter $\theta$.\par
    
    \section{Methodology Implementation} \label{ch 6}
        In this section, we implement the pairs trading strategy in three different scenarios. We explain the first scenario entirely, and then we discuss specific differences between the two other scenarios. \par 
        In the first scenario, we only trade a particular pair of coins every week, but the coins can change over the upcoming weeks. In other words, the optimal pair is selected and traded for one week, but the pair can be replaced with other coins in the upcoming weeks. In the formation period, we define the pair spread value with a constant intercept as follows
        \begin{equation}\label{eq 4.29}
            S_t = P^1_t - \beta P^2_t -\alpha
	    \end{equation}
        where $P^i_t $ is the price of asset i at time t, $\alpha$ is a constant, and the coefficient  $\beta$ tells us the number of shares of the second coin to be added to make a mean-reverting spread. Using ordinary least squares regression, we estimate $\alpha$ and $\beta$. Then, for each pair of coins, the Engle-Granger method and the KSS cointegration test are implemented to test linear cointegration and nonlinear cointegration respectively. Here, the linear unit-root test is the augmented Dickey-Fuller and nonlinear unit-root test is KSS threshold unit-root test.  The confidence level in both tests is 1\%.\par
\begin{figure}[H]
    	    \includegraphics[scale=0.62]{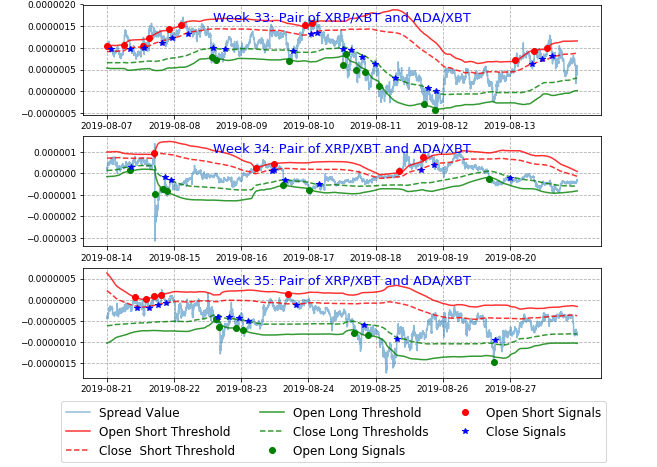}
    	    \centering
    	    \caption {Pairs trading strategy backtesting during some trading weeks}
    	    \label{fig 6.1}
        \end{figure}

        In the next step, by assuming that the spread of cointegrated pairs follows the Ornstein-Uhlenbeck process, we calibrate the parameter $\theta$ in the equation \ref{eq 4.19} and using the equation \ref{eq 4.26}, we estimate the corresponding half-life. Usually, there exist several potential linear and nonlinear cointegrated pairs. Among them, we choose the pair with the least half-life duration. In order to make the spread scalable, we  normalize the spread by calculating  Z-score as follows
        \begin{equation}\label{eq 4.30}
                Z^{score}_t=\frac{S_t-\overline{S_t}}{\hat{\sigma}(S_t)}=\frac{S_i-\sum_{i=t-N+1}^{t}\left({S_i}/{N}\right)}{\sqrt{\sum_{i=t-N+1}^{t}\left(S_i-\sum_{i=t-N+1}^{t}\left({S_i}/{N}\right)\right)^2/(N-1)}}
        \end{equation}
        where, $\bar{S_t}$ is rolling mean, $\hat{\sigma}(S_t)$ is rolling standard deviation, and $N$ is look-back window. \textcite{chan2013algorithmic} formulated the half-life of mean reversion as a look-back window to find rolling mean and rolling standard deviation. We implement a heuristic method to determine the look-back window of the Z-score. First, we apply the estimated half-life of the Ornstein-Uhlenbeck process as a half-life of an exponential moving average (EMA). The EMA for a series S can be calculated recursively as follows
        \begin{equation}\label{eq 4.31}
	        Z_{t}={\begin{cases}S_{1},&t=1\\\lambda \cdot S_{t}+(1-\lambda )\cdot Z_{t-1},&t>1\end{cases}}
	    \end{equation} 
	    where the coefficient $\lambda$ represents the degree of weighting decrease, a constant smoothing factor between 0 and 1. A higher $\lambda$ can discount older observations faster. By expanding out the equation \ref{eq 4.31} we can obtain
	    \begin{equation}\label{eq 4.32}
            {{\text{EMA}}_{\text{Now}}={\lambda \left[s_{1}+(1-\lambda )s_{2}+(1-\lambda )^{2}s_{3}+(1-\lambda )^{3}s_{4}+\cdots \right]}}
        \end{equation}
        where $s_{1}$ is the spread value at the moment, $s_{2}$ is the spread value of one minute before, and so on. If we limit the number of terms and omit the terms after k terms, then The weight omitted by stopping after t terms is $\lambda (1-\lambda )^{t}\left[1+(1-\lambda )+(1-\lambda )^{2}+\cdots \right]$, and consequently, the weight omitted by stopping after k terms over the total terms is equal to:
        \begin{equation}\label{eq 4.33}
            \text{fraction} =\frac{\lambda (1-\lambda )^{t}\left[1+(1-\lambda )+(1-\lambda )^{2}+\cdots \right]}{\lambda \left[1+(1-\lambda )+(1-\lambda )^{2}+\cdots \right]}=\left(1-\lambda \right)^t
        \end{equation}
        To have half of the weights, we set the above fraction to 0.5, and  obtain   
        \begin{equation}\label{eq 4.34}
            t_{1/2} = \frac{\ln(0.5)}{\ln(1-\lambda)}
        \end{equation}
        Since $\lambda \to 0$ as $N \to \infty$, we know $\ln \,(1-\lambda )$ approaches $-\lambda$ as $N$ increases. This gives:
        \begin{equation}\label{eq 4.35}
            t_{1/2} \approx \frac{\ln(0.5)}{-\lambda} = \frac{\ln(2)}{\lambda}
        \end{equation}
        Using equations \ref{eq 4.26}, and \ref{eq 4.35} we deduce that $\lambda=\hat{\theta}$. Now, we find look-back window of simple moving average, $N_{SMA}$ with respect to $\lambda_{EMA}$. When both simple moving average and exponential moving average have the same center of mass, $N_{SMA}$ can be written as follows
        \begin{equation}\label{eq 4.36}
            \lambda_{EMA} =\frac{2}{N_{SMA}+1},\quad \Rightarrow \quad N_{SMA} = \frac{2}{\lambda_{EMA}}-1
        \end{equation}
        We use $N_{SMA}$ as the rolling window of moving average and moving standard deviation in the Z-score formula in the equation \ref{eq 4.30}. Consequently, the threshold for opening a position and unwinding a position is set as follows. If the Z-score diverges two historical standard deviations, then a position is opened. If the Z-score converges back and drops below one historical standard deviation, then the position is closed. Thus, Pairs trading signals can be followed by \\
        \begin{tcolorbox}
            \center{
                if $ Z^{score}_{t-1} > -2$, and $Z^{score}_{t-2}<-2 \Rightarrow$ enter long signal at time $t$,\\
                if $ Z^{score}_{t-1} < -1$, and $Z^{score}_{t-2}>-1 \Rightarrow$ exit long signal at time $t$,\\
	            if $ Z^{score}_{t-1} < +2$, and $Z^{score}_{t-2}>+2 \Rightarrow$ enter short signal at time $t$,\\
	            if $ Z^{score}_{t-1} > +1$, and $Z^{score}_{t-2}< +1 \Rightarrow$ exit short signal at time $t$.
	            }
	    \end{tcolorbox}
        In the trading period, Since we cannot trade a fraction of coins, we should reconstruct the coefficients of the equation \ref{eq 4.28} to make the spread realistic for trading. If $|{\beta}| < 1$, we divide the spread to  $|\beta|$ and then round the its coefficients. Then the modified Spread is 
        \begin{equation}\label{eq 4.37}
            S_t^{modified} = w_1P^1_t +w_2 P^2_t +c
        \end{equation}
        where $w_1$ and $w_2$ are integers. Now we execute the orders. at time t, we calculate the mid price of the assets. Mid price is the average between the best bid and the best ask. Best bid is the most high price offered by buyers and best ask is the lowest price offered by sellers. Knowing the minimum price increment which is the minimum difference between price levels at which a contract can trade, we calculate the feasible best bid and best ask and update them at each moment. Whenever we have trading signals, we update best bid and best ask price for the corresponding coin. In order to backtest of an order execution we need to know the market price of asset at each time. Figure \ref{fig 6.1} demonstrates the backtesting of the strategy in three arbitrary weeks among thirty-nine trading weeks.\par
    	In the second scenario, instead of choosing a pair of coins at the formation period, we build the spread formula by combining a portfolio of coins. Consequently, we have a pair of baskets of coins available in Bitmex exchange. In this case, long and short positions consist of a portfolio of cryptocurrencies. The coin's weights and positions in our portfolio are the only items that can be changed every trading week. Note that the coins with the US dollar base are denominated to the XBT base. Moreover, due to a pretty high correlation between XBT perpetual and XBT futures, it does not make sense to use them simultaneously, so we only work with XBT perpetual. The spread is defined as follows
    	\begin{equation}\label{eq 4.38}
    	    S_t = w_1P^1_t+w_2P^2_t+\cdots+w_9P^9_t
    	\end{equation}
        where $w_i$ are integers and $min(|w_i|)=1$. In this scenario, The Johansen test is used to find the coefficient of multiple coins. Indeed, some coins have positive coefficients, and others have negative coefficients. Like the first scenarios, we have thisty nine trading weeks. The only difference is that we trade all coins every week. Most of the time, the Johansen test gives us plenty of coins combination to make the stationary spread. Our priority in choosing the optimal combination is the minimum half-life of the spreads. The other steps of formation and trading period are the same as the first scenario. \par
        Finally, in the third scenario, by ignoring the pair's selection step, we assume that we know which pair is the best for trading to observe the outcome of trading every possible pair of cryptocurrency coins during the whole thirty-nine trading weeks without changing the items. This scenario is similar to the second one, except we trade each pair in a year without finding the optimal assets. Here, the moving window is set to be one day, so we do not need to find the mean-reverting process's half-life. Note that this scenario is not a comprehensive strategy, though it gives us an insight into an arbitrary coins selection step.
    \section{Empirical Results}\label{ch 7}
        As we discussed earlier, P\&L calculation in BitMEX is based on the XBT. When a contract is quoted in USD, we have to find a multiplier coefficient to denominate the coin to XBT. Most of the futures contracts are denominated in XBT. Thus their multiplier is equal to $1$. A Quanto ETH/USD contract has a fixed multiplier, which is $0.000001$ XBT. The coin XBTUSD is an inverse contract, which means by shorting it, we long US dollars, and by long the contract, we short US dollars. In this case, the P\&L should be calculated conversely, which means that bitcoin's entry price is the exit price of the US dollar, and the exit price of XBT is the entry price of the US dollar. \par
\begin{figure}[H]
			\includegraphics[scale=0.65]{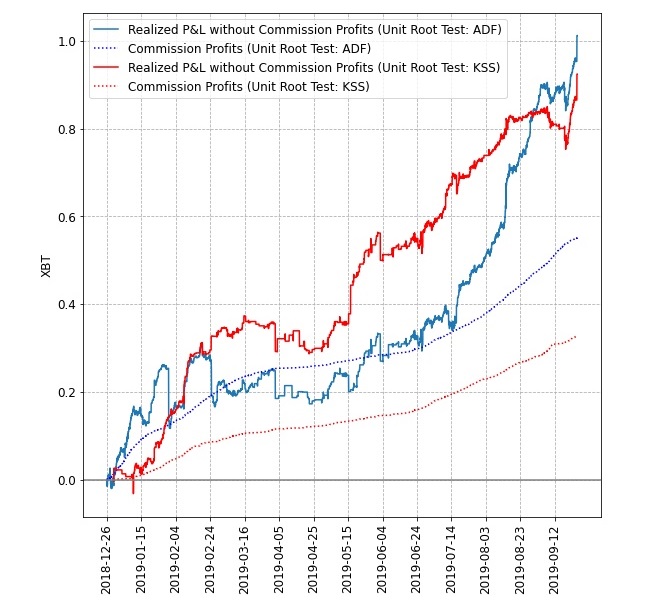}
			\centering
			\caption {Pairs trading strategy P\&L in the first scenario}
			\label{fig 7.1}
		\end{figure}

            Figure \ref{fig 7.1} shows realized profit and loss results in scenario 1. Profit and loss are realized when a contract that we open is closed. On the other hand, unrealized P\&L refers to profits or losses on paper, but the relevant transactions have not been completed. In general, unrealized P\&L has higher fluctuation, and it is useful to calculate the maximum drawdown of the strategy, which is an indicator of downside risk. Realized P\&L consists of commission fees (or commission profits) and a difference between the price of entering and exiting a position. In this scenario, we test two separate unit root tests that result different candidate pairs and different P\&L. Table \ref{table 4} demonstrates the unit-root test results for the chosen pairs in each week. Note that unit root tests are required to find cointegrated coins, but they are not sufficient to find the most profitable trades. The results show that during the trading period, some coins are more liquid. Since the market liquidity is a key factor of a strategy's profitability, using a nonlinear unit root test such as the KSS test does not necessarily increase the profitability of the strategy.\par
\begin{figure}[H]
			\includegraphics[scale=0.65]{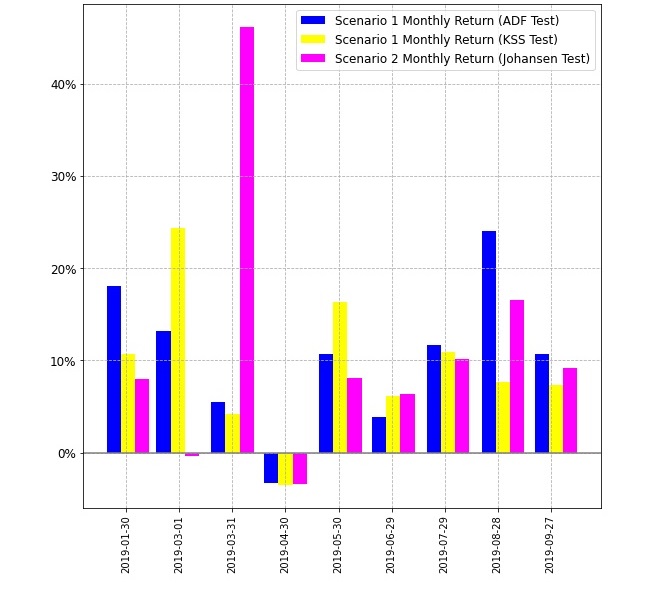}
			\centering
			\caption {Monthly return in the first and second scenarios}
			\label{fig 7.2}
		\end{figure}

		This strategy's initial investment cannot be fixed over time because each week, we have a different combination of assets. However, every week, the coins' weights are set so that the initial capital is around $1$ XBT. Figure \ref{fig 7.2} demonstrates the monthly return of pairs trading strategy for the first and the second scenarios, during 9 months. As is shown, around $90\%$ of the months have positive returns. The average monthly returns of the strategy under ADF and KSS tests are equal to $17.3\%$ and 13.9\% respectively. Commission profits make $35\%$ and $26\%$ of the total profit under ADF and KSS tests respectively. The maximum drawdown is approximately $0.15$ XBT, which has a reasonable amount. \par
\begin{figure}[H]
			\includegraphics[scale=0.65]{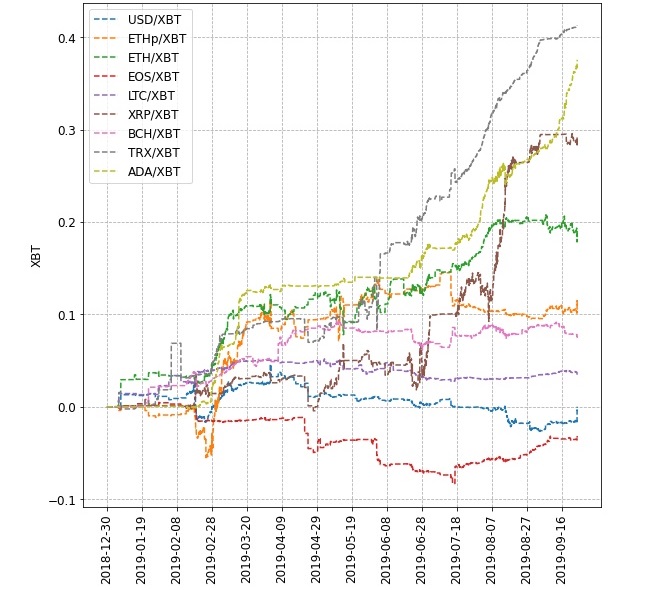}
			\centering
			\caption {P\&L  of the second scenario separately for each coin}
			\label{fig 7.3}
		\end{figure}

		In the second scenario, which includes the basket pairs trading strategy with a dynamic Johansen cointegration test, the total (realized) P\&L is equal to $1.44$ XB, and commission profit is equal to $0.81$ XBT. Figure \ref{fig 7.3} demonstrates the separated profit of each coin during thirty-nine weeks. In table \ref{table 5}, the Johansen test's coefficients of a linear combination of time series to produce a stationary portfolio are shown. Usually, there are multiple available coefficients to produce such a portfolio. We considered three criteria to choose the best one. First, the p-value of the corresponding coefficients should be less than 10\%. Second, at the beginning of each trading week, the unit value of the portfolio should preferably less than 1 XBT (in practice less than 3 XBT). Finally, among our remained choices, the optimal coefficients should have the shortest half-life.\par
		In order to find the risk-adjusted return of the strategy, we identify the Shape ratio, which is calculated by subtracting the risk-free rate from the return of the strategy's P\&L and dividing that result by the standard deviation of the strategy's P\&L. We assume that the risk-free rate is equal to 
		zero. The standard deviation of this strategy is equal to $24.7\%$. Therefore, the Shape ratio is equal to $7.94$, which is considerable. We should also mention that the standard deviations of the first scenario under ADF and KSS tests are equal to $30.0\%$ and $25.8\%$, and consequently, the Sharpe ratios of the first scenario under ADF and KSS tests are equal to $6.96$ and $6.57$ respectively.\par
        Table \ref{table 3} shows the quantities calculated for the third scenario's result. The initial investment for all separate pairs is 1 XBT, which is divided between two coins. Besides, the trading days for all pairs are the same as previous scenarios. In column 7, we introduce return over maximum drawdown. Return over maximum drawdown (RoMaD) is a risk-adjusted return, which is the Sharpe Ratio alternative. We also compare the pairs trading strategy with a buy-and-hold strategy. In the buy-and-hold strategy, we buy relative coins and hold them for the whole trading period despite market fluctuations. \par
        To implement the Buy-and-Hold strategy, we long both cryptocurrency coins with the same weight (half for each coin), and after a year, we short our open positions. The days with profit among separate pairs vary from $63\%$ to $82\%$. Most of the trades take place with limit orders, and a few of them are traded with market orders. In this scenario, P\&L vary from $0.1$ XBT to $3.63$ XBT. In all cases, the Buy-and-Hold strategy has negative returns, and the pairs trading strategy outperforms far better than it. The risk factor of this scenario is more than previous scenarios. According to table \ref{table 6}, we can see a different Sharpe ratio level when we choose different pairs of coins. Concerning the Sharpe ratio levels in the third scenario, we can also recognize that pairs trading strategy in the first two scenarios outperforms more than $70\%$ of single pair of coins in scenario three, which means that basket pairs trading strategy has an acceptable performance. \par
        \begin{table}[H]
            \centering
            \begin{tabular}{@{}llll@{}}
                \toprule
                \textbf{\# Column} & \textbf{Description} &\textbf{\# Column} & \textbf{Description}\\ \midrule
                1 & Profitable Days &8 & Average Daily Return \\
                2 & Number of Trades &9 & Annual Buy\&Hold Return \\
                3 & Number of Market Orders &10 & Annual Buy\&Hold STD \\
                4 & Pair P\&L&11 &Ann. Buy\&Hold Sharpe Ratio \\
                5 & Pair Cumululative Return &12 & Annual Pairs Return  \\
                6 & Maximum Drawdown &13 & Annual Pairs STD \\
                7 & Return Over Max. Drawdown &14 & Annual Pairs Sharpe Ratio \\
                \bottomrule
            \end{tabular}
            \caption{Quantities of the table \ref{table 6} columns}
            \label{table 3}
        \end{table}
        \newpage
        \thispagestyle{empty}
        \begin{table}[H]
            \begin{tabular}{@{}cllllll@{}}
                \toprule
                \textbf{}& \multicolumn{3}{c}{\textbf{ADF Unit-Root Test Result}}       & \multicolumn{3}{c}{\textbf{KSS Unit-Root Test Result}}       \\ \midrule
                \multicolumn{1}{c}{\textbf{Week}} & \multicolumn{1}{c}{\textbf{Selected Pair}} & \multicolumn{1}{c}{\textbf{ Statistic}} & \multicolumn{1}{c}{\textbf{P-Value}} & \multicolumn{1}{c}{\textbf{Selected Pair}} & \multicolumn{1}{c}{\textbf{ Statistic}} & \multicolumn{1}{c}{\textbf{P-Value}} \\\midrule
                1& BCH - ADA & -2.58& 0.96\%   & ETHp - BCH& -4.80& 0.00\%   \\
                2& LTC - ADA & -2.98& 0.28\%   & ETHp - BCH& -4.79& 0.00\%   \\
                3& USD - BCH & -3.13& 0.17\%   & LTC - ADA & -2.67& 0.77\%   \\
                4& USD - BCH & -3.20& 0.14\%   & LTC - ADA & -2.63& 0.86\%   \\
                5& LTC - ADA & -2.58& 0.95\%   & LTC - ADA & -2.64& 0.83\%   \\
                6& XRP - BCH & -3.33& 0.09\%   & LTC - ADA & -2.71& 0.68\%   \\
                7& XRP - BCH & -3.07& 0.21\%   & XRP - BCH & -3.40& 0.07\%   \\
                8& XRP - BCH & -4.35& 0.00\%   & XRP - BCH & -4.42& 0.00\%   \\
                9& USD - BCH & -4.72& 0.00\%   & USD - BCH & -2.80& 0.51\%   \\
                10    & USD - BCH & -3.72& 0.02\%   & ETHp - ETH& -3.82& 0.01\%   \\
                11    & ETHp - ETH& -3.43& 0.06\%   & ETHp - ETH& -3.61& 0.03\%   \\
                12    & LTC - XRP & -3.54& 0.04\%   & ETHp - ETH& -3.37& 0.08\%   \\
                13    & BCH - TRX & -3.27& 0.11\%   & USD - TRX & -7.19& 0.00\%   \\
                14    & ETHp - ETH& -3.29& 0.10\%   & ETHp - ETH& -2.96& 0.31\%   \\
                15    & LTC - XRP & -4.07& 0.01\%   & ETHp - TRX& -4.32& 0.00\%   \\
                16    & LTC - XRP & -4.91& 0.00\%   & USD - XRP & -3.07& 0.22\%   \\
                17    & EOS - TRX & -4.04& 0.01\%   & USD - TRX & -3.91& 0.01\%   \\
                18    & XRP - TRX & -3.09& 0.20\%   & XRP - TRX & -10.55& 0.00\%   \\
                19    & XRP - TRX & -3.56& 0.04\%   & XRP - TRX & -10.69& 0.00\%   \\
                20    & XRP - TRX & -3.08& 0.20\%   & XRP - TRX & -2.80& 0.52\%   \\
                21    & ETH - TRX & -5.31& 0.00\%   & XRP - TRX & -3.06& 0.23\%   \\
                22    & XRP - TRX & -2.89& 0.38\%   & XRP - TRX & -3.72& 0.02\%   \\
                23    & XRP - TRX & -2.82& 0.47\%   & XRP - TRX & -3.91& 0.01\%   \\
                24    & EOS - LTC & -3.44& 0.06\%   & XRP - TRX & -4.09& 0.00\%   \\
                25    & XRP - TRX & -3.10& 0.19\%   & XRP - TRX & -4.67& 0.00\%   \\
                26    & XRP - TRX & -3.50& 0.05\%   & XRP - TRX & -4.69& 0.00\%   \\
                27    & XRP - ADA & -4.17& 0.00\%   & XRP - ADA & -3.69& 0.02\%   \\
                28    & XRP - ADA & -3.75& 0.02\%   & XRP - ADA & -4.81& 0.00\%   \\
                29    & EOS - XRP & -3.65& 0.03\%   & EOS - XRP & -2.78& 0.55\%   \\
                30    & XRP - ADA & -3.54& 0.04\%   & XRP - ADA & -3.93& 0.01\%   \\
                31    & XRP - ADA & -4.17& 0.00\%   & XRP - ADA & -3.83& 0.01\%   \\
                32    & XRP - ADA & -4.25& 0.00\%   & XRP - ADA & -3.73& 0.02\%   \\
                33    & XRP - ADA & -4.16& 0.00\%   & XRP - ADA & -3.05& 0.23\%   \\
                34    & XRP - ADA & -3.62& 0.03\%   & XRP - BCH & -5.20& 0.00\%   \\
                35    & XRP - ADA & -3.39& 0.07\%   & XRP - BCH & -3.37& 0.08\%   \\
                36    & XRP - ADA & -3.47& 0.05\%   & XRP - BCH & -3.33& 0.09\%   \\
                37    & EOS - XRP & -4.19& 0.00\%   & EOS - XRP & -2.94& 0.33\%   \\
                38    & XRP - ADA & -4.26& 0.00\%   & XRP - ADA & -4.24& 0.00\%   \\
                39    & XRP - ADA & -3.62& 0.03\%   & XRP - ADA & -4.74& 0.00\%   \\ \bottomrule
            \end{tabular}
            \caption{Scenario 1 unit-root test results for the selected pairs (p-values $<$ 1\%)}
            \label{table 4}
        \end{table}
        \newpage
        \thispagestyle{empty}
        \begin{table}[H]
            \begin{tabular}{@{}rrrrrrrrrrr@{}}
                \toprule
                \multicolumn{1}{c}{\textbf{\footnotesize W}} & \multicolumn{1}{c}{\textbf{\footnotesize USD}} & \multicolumn{1}{c}{\textbf{\footnotesize ETHp}} & \multicolumn{1}{c}{\textbf{\footnotesize ETH}} & \multicolumn{1}{c}{\textbf{\footnotesize EOS}} & \multicolumn{1}{c}{\textbf{\footnotesize LTC}} & \multicolumn{1}{c}{\textbf{\footnotesize XRP}} & \multicolumn{1}{c}{\textbf{\footnotesize BCH}} & \multicolumn{1}{c}{\textbf{\footnotesize TRX}} & \multicolumn{1}{c}{\textbf{\footnotesize ADA}} & \multicolumn{1}{c}{\textbf{\footnotesize P-Value}} \\ \midrule
                1      & -801     & -1668   & 3       & -54  & 35   & -5   & 1    & 8392 & 957  & 5.7\%   \\
                2      & 950      & 1192    & -13     & 248  & -51  & -324 & 1    & 49336& 941  & 0.1\%   \\
                3      & 101      & -191    & 3       & -89  & -5   & 91   & 1    & -11239& -109 & 7.2\%   \\
                4      & 39       & -437    & 3       & -139 & 1    & 101  & 2    & -9902& 132  & 3.9\%   \\
                5      & -849     & -927    & 5       & -153 & 62   & 35   & -1   & -31922& -1174& 1.6\%   \\
                6      & -744     & -765    & 3       & -102 & 53   & 10   & -1   & -23392& -857 & 0.4\%   \\
                7      & 1781     & 2139    & -10     & 365  & -129 & -107 & 1    & 64388& 2648 & 0.4\%   \\
                8      & 902      & 707     & 1       & 35   & -5   & -3906& 2    & -3663& -3688& 0.0\%   \\
                9      & -804     & -1462   & 4       & 9    & 4    & 3112 & -1   & 6004 & -5639& 0.8\%   \\
                10     & 995      & 5682    & -18     & -42  & -4   & 376  & -1   & 44355& -51231& 0.1\%   \\
                11     & 693      & 1682    & -6      & -14  & -3   & -2127& 1    & 425  & -1843& 6.2\%   \\
                12     & 422      & 2618    & -8      & -71  & 2    & 411  & -1   & 19243& -23357& 4.2\%   \\
                13     & -581     & 884     & -5      & 38   & 1    & 3419 & -4   & 9880 & -1930& 0.3\%   \\
                14     & 1944     & 6168    & -26     & -75  & -1   & -3325& 1    & -3900& -7376& 0.0\%   \\
                15     & -245     & 261     & -1      & -9   & 1    & 1043 & -2   & 3371 & 2522 & 4.6\%   \\
                16     & 681      & 2549    & -10     & -45  & 1    & -909 & -3   & 2482 & 1785 & 0.0\%   \\
                17     & 3541     & 3357    & 5       & -427 & 1    & -9619& 1    & -68588& -98  & 0.0\%   \\
                18     & -708     & -155    & -14     & 350  & -7   & 8104 & 1    & -29634& -5514& 0.6\%   \\
                19     & -413     & 536     & -6      & 107  & -9   & 7203 & 1    & -58912& -196 & 0.0\%   \\
                20     & 89       & -1120   & 8       & -20  & 3    & -3198& 1    & 12322& -2039& 0.0\%   \\
                21     & -364     & 7       & 6       & -118 & 2    & -60  & -1   & -5811& 3299 & 0.0\%   \\
                22     & 230      & 586     & -6      & 1    & 3    & -2038& -1   & 38127& -5   & 0.0\%   \\
                23     & 808      & 1404    & -14     & -215 & 4    & -5523& -1   & 114796& 3169 & 0.0\%   \\
                24     & 270      & 371     & -8      & 40   & -3   & 2774 & 1    & -14789& -42  & 0.2\%   \\
                25     & 532      & 741     & -14     & 14   & -1   & 4153 & 1    & -21697& 616  & 0.0\%   \\
                26     & 270      & 32      & 5       & -56  & -2   & -5327& -1   & 32058& 7101 & 0.0\%   \\
                27     & 143      & 47      & 3       & 27   & -1   & -4177& -1   & 8913 & 8461 & 0.0\%   \\
                28     & 168      & 160     & -4      & 39   & 1    & -134 & 1    & -12331& 1426 & 4.5\%   \\
                29     & 280      & 220     & -1      & 193  & 2    & -129 & -2   & -52784& 7691 & 3.3\%   \\
                30     & 40       & 235     & -11     & -187 & 3    & 16945& -1   & -31205& -27725& 0.0\%   \\
                31     & 56       & -106    & 4       & 44   & -1   & -7622& 1    & 18927& 11356& 0.2\%   \\
                32     & 251      & -139    & 7       & 78   & -1   & -13125& 2    & 24497& 18847& 0.0\%   \\
                33     & 631      & -87     & 7       & 121  & -1   & -17469& 3    & 27017& 25808& 0.0\%   \\
                34     & 623      & 105     & -2      & 65   & -1   & -6828& 1    & 10793& 14331& 7.4\%   \\
                35     & -93      & 231     & -10     & -145 & -1   & 159  & 4    & 28646& 4221 & 7.8\%   \\
                36     & 1376     & 116     & -1      & 297  & 4    & -20457& 3    & 53880& 18374& 0.0\%   \\
                37     & -326     & -211    & 5       & -46  & 1    & -83  & 1    & -1369& -6858& 4.9\%   \\
                38     & -214     & -179    & 3       & 5    & -1   & -35  & 1    & 7432 & -8096& 0.4\%   \\
                39     & 1297     & 614     & -15     & 129  & 7    & -8576& -1   & 12724& 34950& 0.2\%   \\ \bottomrule
            \end{tabular}
            \caption{Johansen test's coefficients to produce a stationary portfolio (p-values $<$ 10\%)}
            \label{table 5}
        \end{table}
        \begin{landscape}
            \thispagestyle{empty}
            \begin{table}[H]
                \centering
                \resizebox{1.41\textwidth}{!}{%
                \begin{tabular}{@{}lllllllllllllll@{}}
                    \toprule
                    \textbf{Pairs}   & \textbf{(1)} & \textbf{(2)} & \textbf{(3)} & \textbf{(4)} & \textbf{(5)} & \textbf{(6)} & \textbf{(7)} & \textbf{(8)} & \textbf{(9)} & \textbf{(10)} & \textbf{(11)} & \textbf{(12)} & \textbf{(13)} & \textbf{(14)} \\ \midrule
                    ETH(P)-ETH(F)   & 64.38\% & 3262 & 369 & 0.1 XBT & 10.17\% & -40.41\% & 0.25 & 0.03\% & -30.24\% & 66.23\% & -0.46 & 10.17\% & 34.72\% & 0.29 \\
                    ETH(P)-EOS   & 63.56\% & 2087 & 119 & 0.1 XBT & 10.17\% & -40.41\% & 0.25 & 0.03\% & -41.23\% & 74.36\% & -0.55 & 29.71\% & 33.59\% & 0.88 \\
                    ETH(P)-XRP   & 63.01\% & 1994 & 51 & 0.37 XBT & 36.58\% & -45.32\% & 0.81 & 0.08\% & -42.64\% & 71.57\% & -0.60 & 36.58\% & 33.40\% & 1.10 \\
                    ETH(P)-LTC   & 63.56\% & 2229 & 99 & 0.89 XBT & 88.88\% & -25.91\% & 3.43 & 0.17\% & -25.35\% & 73.87\% & -0.34 & 88.88\% & 29.48\% & 3.01 \\
                    ETH(F)-LTC   & 68.77\% & 1853 & 169 & 0.84 XBT & 84.30\% & -13.01\% & 6.48 & 0.17\% & -31.91\% & 49.99\% & -0.64 & 84.30\% & 27.08\% & 3.11 \\
                    ETH(F)-EOS   & 69.95\% & 2082 & 236 & 0.87 XBT & 87.08\% & -11.74\% & 7.42 & 0.17\% & -48.81\% & 50.99\% & -0.96 & 86.84\% & 25.32\% & 3.43 \\
                    EOS-XRP   & 67.40\% & 2000 & 126 & 0.93 XBT & 93.12\% & -15.47\% & 6.02 & 0.18\% & -62.88\% & 57.89\% & -1.09 & 93.12\% & 26.75\% & 3.48 \\
                    ETH(F)-XRP   & 71.31\% & 1997 & 129 & 1.04 XBT & 104.21\% & -12.15\% & 8.58 & 0.19\% & -50.22\% & 46.73\% & -1.07 & 103.93\% & 24.99\% & 4.16 \\
                    LTC-XRP   & 67.58\% & 1850 & 96 & 0.99 XBT & 98.83\% & -22.32\% & 4.43 & 0.19\% & -48.42\% & 56.67\% & -0.85 & 99.10\% & 23.57\% & 4.20 \\
                    ETH(F)-BCH   & 73.22\% & 2056 & 177 & 1.46 XBT & 145.57\% & -20.95\% & 6.95 & 0.24\% & -52.16\% & 83.16\% & -0.63 & 145.17\% & 32.61\% & 4.45 \\
                    ETH(P)-BCH   & 69.86\% & 2284 & 121 & 1.4 XBT & 139.56\% & -28.57\% & 4.88 & 0.24\% & -44.82\% & 99.21\% & -0.45 & 139.56\% & 30.26\% & 4.61 \\
                    EOS-BCH   & 69.95\% & 2069 & 158 & 1.57 XBT & 156.81\% & -19.43\% & 8.07 & 0.26\% & -63.21\% & 89.91\% & -0.70 & 156.38\% & 32.40\% & 4.83 \\
                    ETH(P)-ADA   & 68.77\% & 2395 & 132 & 1.18 XBT & 118.09\% & -42.14\% & 2.80 & 0.21\% & -41.81\% & 77.97\% & -0.54 & 118.09\% & 22.83\% & 5.17 \\
                    BCH-TRX   & 74.04\% & 2126 & 213 & 1.67 XBT & 166.58\% & -24.16\% & 6.90 & 0.26\% & -58.02\% & 120.84\% & -0.48 & 166.13\% & 31.80\% & 5.22 \\
                    LTC-BCH   & 72.60\% & 2067 & 180 & 1.64 XBT & 163.96\% & -19.18\% & 8.55 & 0.27\% & -50.32\% & 89.16\% & -0.56 & 163.96\% & 31.43\% & 5.22 \\
                    ETH(P)-TRX   & 69.59\% & 2292 & 152 & 1.53 XBT & 152.54\% & -23.74\% & 6.43 & 0.25\% & -35.92\% & 109.73\% & -0.33 & 152.54\% & 29.24\% & 5.22 \\
                    XRP-BCH   & 73.50\% & 2237 & 132 & 1.89 XBT & 189.07\% & -27.33\% & 6.92 & 0.29\% & -64.64\% & 87.70\% & -0.74 & 188.55\% & 33.29\% & 5.66 \\
                    ETH(F)-TRX   & 73.22\% & 2199 & 260 & 1.67 XBT & 167.43\% & -24.25\% & 6.90 & 0.27\% & -43.49\% & 95.50\% & -0.46 & 166.98\% & 27.09\% & 6.16 \\
                    EOS-TRX   & 72.40\% & 2277 & 302 & 1.72 XBT & 171.85\% & -24.80\% & 6.93 & 0.27\% & -54.48\% & 101.52\% & -0.54 & 171.38\% & 27.39\% & 6.26 \\
                    EOS-LTC   & 71.51\% & 2223 & 216 & 1.63 XBT & 162.72\% & -10.87\% & 14.97 & 0.26\% & -43.17\% & 60.63\% & -0.71 & 162.72\% & 22.03\% & 7.39 \\
                    XRP-TRX   & 73.50\% & 2435 & 269 & 1.99 XBT & 198.86\% & -26.81\% & 7.42 & 0.30\% & -55.89\% & 99.29\% & -0.56 & 198.32\% & 24.60\% & 8.06 \\
                    EOS-ADA   & 74.32\% & 2294 & 173 & 1.79 XBT & 178.89\% & -12.46\% & 14.36 & 0.28\% & -60.37\% & 65.43\% & -0.92 & 178.40\% & 21.62\% & 8.25 \\
                    LTC-TRX   & 74.79\% & 2219 & 239 & 2.1 XBT & 210.23\% & -24.54\% & 8.57 & 0.31\%  & -38.41\% & 100.79\% & -0.38 & 210.23\% & 24.40\% & 8.62 \\
                    ETH(F)-ADA   & 75.14\% & 2244 & 203 & 1.89 XBT & 189.13\% & -11.46\% & 16.50 & 0.29\% & -49.38\% & 55.99\% & -0.88 & 188.61\% & 18.85\% & 10.00 \\
                    BCH-ADA   & 79.23\% & 2390 & 164 & 2.84 XBT & 283.66\% & -20.57\% & 13.79 & 0.37\% & -63.68\% & 92.75\% & -0.69 & 282.88\% & 27.72\% & 10.21 \\
                    XRP-ADA   & 75.41\% & 2317 & 153 & 2.23 XBT & 223.01\% & -13.00\% & 17.15 & 0.32\% & -61.78\% & 62.26\% & -0.99 & 222.40\% & 20.11\% & 11.06 \\
                    LTC-ADA   & 72.95\% & 2181 & 145 & 2.23 XBT & 222.68\% & -11.02\% & 20.20 & 0.32\% & -43.15\% & 64.68\% & -0.67 & 222.07\% & 17.87\% & 12.43 \\
                    TRX-ADA   & 81.69\% & 3122 & 479 & 3.63 XBT & 363.17\% & -24.12\% & 15.06 & 0.41\% & -55.05\% & 103.88\% & -0.53 & 362.18\% & 17.77\% & 20.38\\ \bottomrule
                \end{tabular}}
                \caption{The third scenario empirical results}
                \label{table 6}
            \end{table}
            \raisebox{0 em}{\makebox[\linewidth]{\thepage}}
        \end{landscape}
    \section{Conclusion}\label{ch 8}
        In this paper, we achieved a dynamic cointegrated-based pairs trading strategy in three different scenarios. The first and second scenarios incorporated the whole pairs trading strategy steps from the formation period to the trading period. Between these two scenarios, the second scenario has the Sharpe ratio of 7.94 outperforms the first scenario with the Sharpe ratios of $6.96$ and $6.57$ under ADF and KSS tests. So, applying a portfolio of assets to build the spread model improves the risk-adjusted profitability of the pairs trading strategy. The dynamic pairs trading strategy's maximum drawdown was not high, which means that the risk of deploying this strategy is satisfactory. Note that if the strategy's maximum drawdown level were significant, we would have executed the orders with stop loss or trailing stop loss. Moreover, we can perceive that by increasing the number of coins and diversification of coins' collection, we can lower the risk of trading at a satisfactory level. Likewise, the strategy's return, in this case, would be decreased. \par
        The third scenario reveals that pairs selection (the formation period) can dramatically affect the strategy's performance. For instance, the Sharpe ratio of trading ETH futures with ETH perpetual swaps is 0.29, which is low. Conversely, the Sharpe ratio of Cardano-Tron futures exceeded 20, which is considerably high. Although each pair's return or Sharpe ratio is dramatically different in the third scenario, the pairs trading strategy exceeds the naive Buy-and-Hold strategy in the Bitmex exchange. The research's finding shows that some coins such as Tron (TRX), Cardano (ADA), and Ripple (XRP) are desirable candidates to be applied in pairs trading strategy because, in different scenarios, they reveal better performance. Hence it appears that some class of coins has a better potential concerning statistical arbitrage opportunities than the others. \par
        This research has some noticeable advantages, making it stand out from similar studies in the cryptocurrency market. First is the accuracy of data in which minute-binned data create the signals in the formation period.  Besides, to backtest the strategy during the trading period, we simulate the trading signals using best bid/ask quotes and market trades. Similar researches in this area have daily-based data. Furthermore, the cryptocurrency market that we studied is a developing market with more potential for statistical arbitrage opportunities than traditional markets such as the stock market. Finally, we considered almost all limitations in the market microstructure. The market microstructure has a notable effect on the profitability of each strategy. Most of the studies in this domain are not practical since they usually simplify market microstructure, which unrealistically affects the profit. One of the main constraints of market microstructure is volume. It is questionable whether the desired coin is always available at reasonable costs and in reasonable quantities. In this paper, We exclusively take the order execution into account when the asset size is already available at its quoted price (with one or more period gaps after signal generation). This action makes the backtesting much more realistic. Another critical limitation is that some cryptocurrency exchanges do not allow short selling, but we chose an exchange to overcome this limitation. \par
        There are some ideas to extend this paper framework. One could combine this strategy with machine learning tools to predict the jumps in the market and recalculate parameters after each jump. In practice, jumps are the main reason for non-proper signals in the strategy. Furthermore, one can expand stochastic spread using Markov switching models or Bayesian statistics to find optimal trading signals. Finally, considering the cryptocurrency market is expanding quickly,  performing the other studies over other exchanges and other coins might give different results.
    \printbibliography
\end{document}